# Observation of anomalous decoherence effect in a quantum bath at room temperature

Pu Huang[1, #], Xi Kong[1, #], Nan Zhao[2, #], Fazhan Shi[1], Pengfei Wang[1], Xing Rong[1], Ren-Bao Liu[2,*], and Jiangfeng Du[1,*]

1. *Hefei National Laboratory for Physical Sciences at the Microscale and Department of Modern Physics, University of Science and Technology of China, Hefei, Anhui 230026, China*
2. *Department of Physics and Centre for Quantum Coherence, The Chinese University of Hong Kong, Shatin, New Territories, Hong Kong, China*

[#] *These authors contribute equally*
[*] *Corresponding authors*

**Decoherence of quantum objects is critical to modern quantum sciences and technologies. It is generally believed that stronger noises cause faster decoherence. Strikingly, recent theoretical research discovers the opposite case for spins in quantum baths. Here we report experimental observation of the anomalous decoherence effect for the electron spin-1 of a nitrogen-vacancy centre in high-purity diamond at room temperature. We demonstrate that under dynamical decoupling, the double-transition can have longer coherence time than the single-transition, even though the former couples to the nuclear spin bath as twice strongly as the latter does. The excellent agreement between the experimental and the theoretical results confirms the controllability of the weakly coupled nuclear spins in the bath, which is useful in quantum information processing and quantum metrology.**

Coupling with the environment causes decoherence of a quantum object, which is a key issue in quantum sciences and technologies [1-3]. Such coupling is usually understood as classical noises, such as in the spectral diffusion theories, which are widely used in, e.g., magnetic resonance spectroscopy [4, 5] and optical spectroscopy [6, 7]. In modern nanotechnologies and quantum sciences, the relevant environment of a quantum object can be of nanometer or even sub-nanometer size [8-22]. Therefore, the environment itself is of quantum nature. Quantum theories have been developed in recent years to treat the decoherence problem in a quantum bath [23-28]. The quantum theories have been successful in studying decoherence in various systems and predicted some surprising quantum effects [27]. The direct demonstration of the fundamental difference between classical and quantum baths, however, remains elusive.

A recent theoretical study [29] predicted an anomalous decoherence effect (ADE) of a quantum bath on a spin higher than 1/2. The prediction is that the multi-transition can have longer coherence time than the single-transition under dynamical decoupling control, even though the former suffers stronger noises. The nitrogen-vacancy (NV) centre in high-purity diamond, which has an electron spin-1 coupled a nanometer-sized nuclear spin bath, is an ideal system to study the ADE.

The NV centre electron spin (referred to as the centre spin hereafter, see Figure 1a for the structure) has long coherence time (~ms) even at room temperature [21], and is promising for applications in quantum information processing [17-22, 30-35] and nanometrology [36-41]. Besides the applications, the NV centre electron spin system is also a good model system for fundamental research of decoherence [28, 29] and the dynamical decoupling control [42-44]. The decoherence of NV centre electron spins in type-IIa samples is caused by coupling to $^{13}$C nuclear spins within several nanometers from the centre [28, 29], which form a quantum spin bath (see Figure 1c).

Here, we report the experimental observation of the ADE using an NV centre system at room temperature. The observed centre spin decoherence is in excellent agreement with the microscopic theory without fitting parameters. The combined experimental and theoretical results demonstrate the capability of manipulating the evolution of the surrounding $^{13}$C nuclear spins by controlling the centre spin. This manipulation paves the way of exploiting spin baths for quantum information processing [45] and nanometrology [39, 41].

## Results

We demonstrate the ADE using paramagnetic resonance measurement and microscopic calculation. The experiments are based on optically detected magnetic resonance [46] of single NV centres in type-IIa diamond at room temperature (Figure 1b). The calculation based on the quantum many-body theory [26] has no fitting parameter (see Methods).

Under zero field, the centre spin has three eigenstates quantized along the $z$ direction (the NV axis, [111] direction), namely, $|\pm\rangle$, and $|0\rangle$ (see Figure 1d). In the experiment, a weak magnetic field (< 20 Gauss) is applied along the NV axis to split $|+\rangle$ and $|-\rangle$. The triplet spin has both the single-transitions $|0\rangle \leftrightarrow |\pm\rangle$ and the double-transition $|+\rangle \leftrightarrow |-\rangle$ (see Figure 1d). The single-transition coherence $L_{0,+}(t)$ and the double-transition coherence $L_{+,-}(t)$ (see Figure. 1d) are measured for a single NV centre. A local field fluctuation $\hat{b}_z$ will induce transition-energy fluctuations for the double-transition $|+\rangle \leftrightarrow |-\rangle$ as twice strong as for the single-transition $|0\rangle \leftrightarrow |+\rangle$. In this sense, the double-transition is subjected to stronger noises and is expected to have shorter coherence time.

The system has a Hamiltonian $H = H_{\mathrm{NV}} + H_{\mathrm{B}} + H_{\mathrm{hf}}$. The centre spin Hamiltonian is $H_{\mathrm{NV}} = \Delta S_z^2 - \gamma_e B S_z$ with $\Delta$ denoting the zero-field splitting and $\gamma_e$ the electron gyromagnetic ratio. The bath Hamiltonian $H_{\mathrm{B}} = \sum_i \omega I_i^z + \sum_{i,j} \mathbf{I}_i \cdot \vec{\mathbf{D}}_{ij} \cdot \mathbf{I}_j$ contains the nuclear spin Zeeman splitting ($\omega$) and the interaction between nuclear spins ($\vec{\mathbf{D}}_{ij}$). The centre spin couples to the nuclear spins through $H_{\mathrm{hf}} = S_z \sum_j \mathbf{A}_j \cdot \mathbf{I}_j \equiv S_z \hat{b}_z$, where $\mathbf{A}_j$ is the hyperfine coupling to the $j$th nuclear spin $\mathbf{I}_j$. Here the transverse components of the hyperfine coupling have been dropped because they are too weak to cause the centre spin flip. The hyperfine coupling strength depends inverse-cubically on the distance of the nuclear spin from the centre, due to its dipolar form. The relevant bath spins locate within a few nanometres from the centre spin (Figure 1c). The nuclear spins outside this range have too weak hyperfine coupling to contribute to the centre spin decoherence [29]. Thus, within the decoherence timescale (< ms), the centre spin together with ~100 bath spins form a relatively close quantum system.

Viewed from the centre spin, the hyperfine coupling provides a quantum noise field $\hat{b}_z$. Since $\hat{b}_z$ in general do not commute with total Hamiltonian, a certain noise-field eigenstate will evolve to a superposition of different eigenstates of $\hat{b}_z$, leading to quantum fluctuations of the centre spin splitting. The Hamiltonian can also be expressed as

$$H = \sum_{\alpha=0,\pm} |\alpha\rangle\langle\alpha| \otimes \left(\omega_\alpha + H^{(\alpha)}\right), \qquad (1)$$

where $\omega_\alpha = \Delta - \alpha\gamma_e B$ is the eigenenergy of $|\alpha\rangle$, and $H^{(\alpha)} = H_B + \alpha\hat{b}_z$ governs the bath dynamics conditioned on the centre spin state. Now viewed from the bath, the hyperfine coupling is a back action, conditioned on the centre spin state. Thus, the centre spin decoherence is caused by conditional bath evolution, which records the which-way information of the centre spin [25, 29].

Besides the quantum fluctuations, there are also classical thermal fluctuations due to the random orientations of nuclear spins at room temperature. Indeed, the thermal fluctuations (also called inhomogeneous broadening) are much stronger than the quantum fluctuations, and cause the free-induction decay of centre spin coherence within several microseconds. However, the inhomogeneous broadening effect can be totally removed by spin echo [47]. The coexistence of classical and quantum fluctuations and their different effects under spin echo control enable the in-situ test of the classical and quantum theories.

Figures 2a & 2b show the free-induction decay of the centre spin coherence. Both single- and double-transition decoherence have Gaussian decay envelopes $\exp(-t^2/T_2^{*2})$, with the coherence time of the double-transition ($T_2^* = 1.82 \pm 0.08$ µs) about half that of the single-transition ($T_2^* = 3.97 \pm 0.18$ µs). This verifies the scaling relation

$$|L_{+,-}(t)| = |L_{0,+}(t)|^4 \qquad (2)$$

predicted in Ref. [29] for thermal fluctuations. The experimental data are in good agreement with the numerical results obtained by considering the inhomogeneous broadening of a $^{13}$C nuclear spin bath.

The quantum fluctuations become relevant when the inhomogeneous effect is removed by spin echo [47]. Figures 2c & 2d show the Hahn echo signals under an external magnetic field $B = 12.5$ Gauss. The single-transition coherence presents periodic revivals. In contrast, the double-transition coherence decays to zero within several microseconds and does not recover. Such qualitative difference results directly from manipulation of the quantum bath upon the centre spin flip.

Under a weak magnetic field, the centre spin decoherence is mainly induced by the single $^{13}$C nuclear spin dynamics [28, 39]. The dipolar interaction between nuclear spins can be neglected for the moment. Thus, the bath Hamiltonian $H_B$ only contains the nuclear Zeeman energy (with $\gamma_C B/2\pi \sim 13.4$ kHz, $\gamma_C$ being the gyromagnetic ratio of $^{13}$C nuclei). The hyperfine field $\alpha \mathbf{A}_j$ (with $A_j/2\pi \sim 5$ kHz for a nuclear spin $\mathbf{I}_j$ located 1.5 nm from the centre) is comparable to the Zeeman frequency. Each nuclear spin precesses about different local fields $\mathbf{h}_j^{(\alpha)} = -\gamma_C \mathbf{B} + \alpha \mathbf{A}_j$, conditioned on the centre spin state $|\alpha\rangle$. The centre spin decoherence is expressed as [28, 39]

$$L_{\alpha,\alpha'}(t) \approx \prod_j \left|\left\langle I_j^{(\alpha)}(t) \middle| I_j^{(\alpha')}(t) \right\rangle\right|, \qquad (3)$$

where $\left|I_j^{(\alpha)}(t)\right\rangle$ is the precession of the $j$th nuclear spin about the local field $\mathbf{h}_j^{(\alpha)}$ starting from a randomly set initial state $|I_j\rangle$. The conditional evolution of bath spins records the which-way information of the centre spin and causes the decoherence. Upon a flip operation of the centre spin $|\alpha\rangle \leftrightarrow |\alpha'\rangle$, the nuclear spin precession is manipulated as $\left|I_j^{(\alpha)}(t)\right\rangle = e^{-i\mathbf{h}_j^{(\alpha)} \cdot \mathbf{I}_j (t-\tau)} e^{-i\mathbf{h}_j^{(\alpha')} \cdot \mathbf{I}_j \tau} |I_j\rangle$, i.e., the nuclear spin changes its precession direction and frequency. Thus, the nuclear spin bath is manipulated via control of the centre spin. The coherence at the echo time is calculated as [28]

$$L_{\alpha,\alpha'}(2\tau) = \prod_j \left(1 - 2\left|\sin\frac{\mathbf{h}_j^{(\alpha)}\tau}{2} \times \sin\frac{\mathbf{h}_j^{(\alpha')}\tau}{2}\right|^2\right). \qquad (4)$$

When the centre spin is in the state $|0\rangle$, all the nuclear spins precess about the same local field $\mathbf{B}$. This fact leads to the single-transition coherence recovery when the echo time is such that the nuclear spins complete full cycles of precession in a period of free evolution under the magnetic field (i.e., $\gamma_C B \tau = 2n\pi$ for integer $n$). This effect is shown in Figure 2c, which is consistent with previous observations [17]. The height of the recovery peaks decays due to the nuclear-nuclear spin interaction in the bath [28, 39]. For the double-transition coherence, however, the hyperfine couplings are non-zero and therefore the nuclear spins have different local fields for both of the centre spin states $|\pm\rangle$. Consequently, the double-transition coherence has no full recovery under the echo control. Figures 2c & 2d show excellent agreement between the theory and the experimental observation.

A careful examination of the initial decay of the spin-echo signals (for $\gamma_C B \tau < 2\pi$) shows that the scaling relation in equation (2) between the single- and double-transition coherence is approximately satisfied (see Figure 2e). This scaling relation, however, is not the classical noise effect but results from the manipulation of the single nuclear spin dynamics in the short time limit. In the initial spin-echo decay under a weak magnetic field, the short time condition $h_j^{(\alpha)} \tau \ll 1$ is satisfied for most nuclear spins coupled to the centre spin. The short time expansion of equation (4) gives $L_{0,+}(2\tau) \approx \prod_j \left(1 - |\gamma_C \mathbf{B} \times \mathbf{A}_j|^2 \tau^4/8\right)$ and $L_{+,-}(2\tau) \approx \prod_j \left(1 - |2\gamma_C \mathbf{B} \times \mathbf{A}_j|^2 \tau^4/8\right)$, which satisfy the scaling relation in equation (2).

To further explore the quantum nature of the nuclear spin bath, we employ the multi-pulse dynamical decoupling control, to elongate the centre spin coherence time and to make the control effects on the quantum bath more pronounced. To focus on the initial-stage decoherence, we use a weak field ($B$ = 5 Gauss) to have a relatively long time window before the single-transition coherence recovery occurs (about 0.37 ms in Hahn echo). Figure 3 compares the single- and double-transition coherence under the periodic dynamical decoupling (PDD) control by equally spaced sequences of up to five pulses (applied at $\tau, 3\tau, 5\tau\ldots$) [42-44]. With increasing the number of control pulses, the double-transition coherence time increases more than that of the single-transition. Surprisingly, under the five-pulse control, the double-transition has significantly longer coherence time than the single-transition. Such counter-intuitive phenomena unambiguously demonstrate the quantum nature of the nuclear spin bath.

The different dependence on dynamical decoupling of the single- and double-transition decoherence, though counter-intuitive, can be understood with a geometrical picture of nuclear spin precession conditioned on the centre spin state (see Ref. [29] and Figure 4). By repeated flip control $|\alpha\rangle \leftrightarrow |\alpha'\rangle$ of the centre spin, a nuclear spin $\mathbf{I}_j$ precesses alternatively about the local fields $\mathbf{h}_j^{(\alpha)}$ and $\mathbf{h}_j^{(\alpha')}$. The decoherence are caused mainly by the relatively closely located $^{13}$C spins, which have hyperfine fields much stronger than the weak external field ($A_j \gg \gamma_C B$). The local fields $\mathbf{h}_j^{(\pm)} = -\gamma_C \mathbf{B} \pm \mathbf{A}_j$ corresponding to the centre spin states $|\pm\rangle$ are approximately anti-parallel, and the bifurcated nuclear spin precession pathways have small distance $\delta_{+,-}$ at echo time (see Figure 4a). Thus, under dynamical decoupling control of the double-transition, the centre spin decoherence due to the closely located nuclear spins is largely suppressed, which, on the contrary, is not the case for the single transition (see Figure 4b).

Figures 3 shows excellent agreement between the measured centre spin decoherence under PDD and the calculation. Some features of slow oscillations or shoulders in the calculated decoherence do not match those in the measured data, especially for the higher-order dynamical decoupling. This difference is understandable since such features depend sensitively on the specific random positions of a few closely located $^{13}$C spins, which are not determined. In addition, the measured data have some fast fluctuations, especially for small pulse delay time, which are due to the non-ideal pulse control. Nevertheless, the experiments unambiguously confirm the prediction that the double-transition coherence time grows to be longer than that of the single-transitions.

## Discussions

Our observation of the ADE using NV centre coherence establishes the quantum nature of nuclear spins bath at room temperature. Previous studies on coherence control of NV center spins in electron-spin baths [32-34, 42] show that the decoherence there is well described by the classical noise theory. The fundamental difference between nuclear spin baths and electron-spin baths lies in the intra-bath interaction strength relative to the bath-centre spin coupling. For nuclear spin baths, the dipolar interaction between nuclear spins at average distance (~10 Hz) is much weaker than the typical hyperfine coupling (> kHz) [28, 29]. With such weak intra-bath interaction, the diffusion of coherence among nuclear spins is much slower than the decoherence (of a timescale ~ms). Thus, the centre spin and the bath can be regarded as a relatively close quantum system. For electron spin baths, however, the coupling between bath spins at average distance is much stronger than the typical bath-centre coupling. As a result, the coherence will rapidly defuse from closely located bath spins to those at distance during the centre spin decoherence. Therefore, an electron spin bath behaves like a macroscopic open system, and the classical noise theory is valid. For NV centre spin decoherence in electron spin baths, we expect the ADE be absent and the scaling relation in equation (2) be observed instead.

The quantum nature of nuclear spin bath can also be understood by the back-action of the centre spin to the bath. For the transition $|\alpha\rangle \leftrightarrow |\alpha'\rangle$, the Hamiltonian in equation (1) can be expressed in a pseudo-spin form as $H_{\text{ps}} = \frac{1}{2} b_{\alpha\alpha'}^z \sigma_z + H_{\alpha\alpha'}$, where $\sigma_z = |\alpha\rangle\langle\alpha| - |\alpha'\rangle\langle\alpha'|$ is the pseudo-spin operator, $b_{\alpha\alpha'} = H^{(\alpha)} - H^{(\alpha')}$ is the effective noise field to the pseudo-spin, and $H_{\alpha\alpha'} = \frac{1}{2}\left(H^{(\alpha)} + H^{(\alpha')}\right)$ is the effective bath Hamiltonian. For the single-transitions $|0\rangle \leftrightarrow |\pm\rangle$, the effective noise field is $\hat{b}_{0\pm} = \hat{b}_z$, and the double-transition $|+\rangle \leftrightarrow |-\rangle$ has a twice stronger noise as $\hat{b}_{+-} = 2\hat{b}_z$. For the double-transition, the effective bath Hamiltonian $H_{+-} = H_B$, but for the single-transition, the effective bath Hamiltonian $H_{0\pm} = H_B \pm \frac{1}{2}\hat{b}_z$ with an additional term due to the hyperfine coupling, which is the back-action of the centre spin to the bath. For the nuclear spin bath, the hyperfine coupling is typically stronger than the intra-bath interactions, and the back-action strongly modifies the effective bath Hamiltonian. In particular, for this work, the hyperfine coupling provides a much stronger local field than the applied magnetic field for nuclear spins close to the centre spin. In the single-transition case, due to the back-action, the nuclear spins have enhanced precession frequencies in comparison with the double-transition case. Thus, viewed from the centre spin, the effective bath for the double-transition produces noises with lower frequencies than in the single-transition case, and therefore, the centre spin coherence is better protected by the dynamical decoupling control. This explains the ADE observed in nuclear spin baths. In contrast, for electron spin baths, the coupling strength within the bath Hamiltonian $H_B$ is much larger than the back-action term $\frac{1}{2}\hat{b}_z$. For different center spin transitions, the modification of the bath dynamics due to the back-action is negligible. In this sense, the electron spin bath behaves as a classical bath and the ADE should not occur.

Finally, we point out that in this work, the ADE is demonstrated in the weak magnetic field regime (< 20 Gauss), in which the quantum fluctuations is caused mainly by single nuclear spin dynamics. The ADE was predicted in Ref. [29] in the strong field regime where the fluctuations are caused mainly by nuclear spin pair dynamics. These works indicate that the ADE is insensitive to the details of the decoherence mechanisms, but is a universal phenomena due to the quantum nature of the bath.

## Methods

**Experimental setup and measurement methods.** All the experiments are carried out at room temperature. The type-IIa diamond single crystal sample has nitrogen concentration less than 5 ppb and a natural abundance of the $^{13}$C isotope. Individual

NV centres are optically addressed by a confocal microscopy mounted on a piezoelectric scanner, and are identified by the measurement of the anti-bunching effect through the second-order correlation function. The weak magnetic field is generated by three pairs of Helmholtz coils with an accuracy of 1 degree for the direction and 0.01 Gauss for the magnitude. The NV centre is initialized into the $|0\rangle$ state by a 532 nm continuous-wave laser of 10 μs duration. The centre spin is manipulated by resonant microwave pulses. A linear amplifier outputs enough microwave power and a 20 μm diameter copper wire couples the microwave field into the diamond. The double-transition is controlled by composite pulses exciting the single transitions. For a fair comparison, in the single-transition control, pulses with rotation angles $3\pi/2$ and $3\pi$ are used instead of $\pi/2$ and $\pi$, respectively, so that the durations of the control pulses are approximately the same as in the double-transition. To guarantee good frequency selectivity of the microwave pulses, a moderate Rabi frequency (about 5 MHz) is employed. The centre spin state is read out by collecting the fluorescence photons within 420 ns under the 532 nm laser. To build up statistical confidence, we typically repeat each measurement $10^5$ times. All the pulse signals, including the laser, the microwave and the readout triggers, are synchronized by a pulse generator with time resolution of about 3 ns.

**Theoretical model and numerical simulation methods.** In numerical calculation, the nuclear spin bath is generated by randomly placing $^{13}$C atoms on the diamond lattice around the NV centre with a natural abundance 1.1%. Inclusion of about 100 $^{13}$C nuclear spins within 2 nm from the NV centre is sufficient for a converged result of the centre spin decoherence in the timescale considered in this paper. The generated bath does not contain a $^{13}$C in the first few coordinate shells of the NV centre (which has hyperfine coupling >1 MHz), being consistent with the NV centre under the experimental observation. The hyperfine interaction is assumed the dipolar form with the electron spin set at the vacancy site. The spin coherence is calculated by applying the cluster correlation expansion method [26], which takes into account order by order the many-body correlations induced by the dipolar interactions between nuclear spins, and can identify the contribution of each nuclear spin cluster to the total coherence. The converged results are obtained by including clusters containing up to 3 nuclear spins. The microwave pulses are modeled by instantaneous pulses.

**Acknowledgements** This work was supported by National Natural Science Foundation of China under Grant No. 11028510 and 10834005, the Chinese Academy of Sciences, and the National Fundamental Research Program of China under Grant No. 2007CB925200, Hong Kong Research Grant Council/General Research Fund CUHK402410, The Chinese University of Hong Kong Focused Investments Scheme, Hong Kong Research Grant Council HKU10/CRF/08.

**Author Contributions** J.D. designed and supervised the experiments. X.K., F.S., P.W., X.R. and J.D. prepared the experimental setup. P.H. and X.K. performed the experiments**.** R.B.L. and N.Z. formulated the theory. N.Z. carried out the calculation. R.B.L., N.Z. and P. H. wrote the paper. All authors analyzed the data, discussed the results and commented on the manuscript.

**Correspondence** and requests for materials should be addressed to J.D. (djf@ustc.edu.cn) or R.B.L. (rbliu@phy.cuhk.edu.hk).

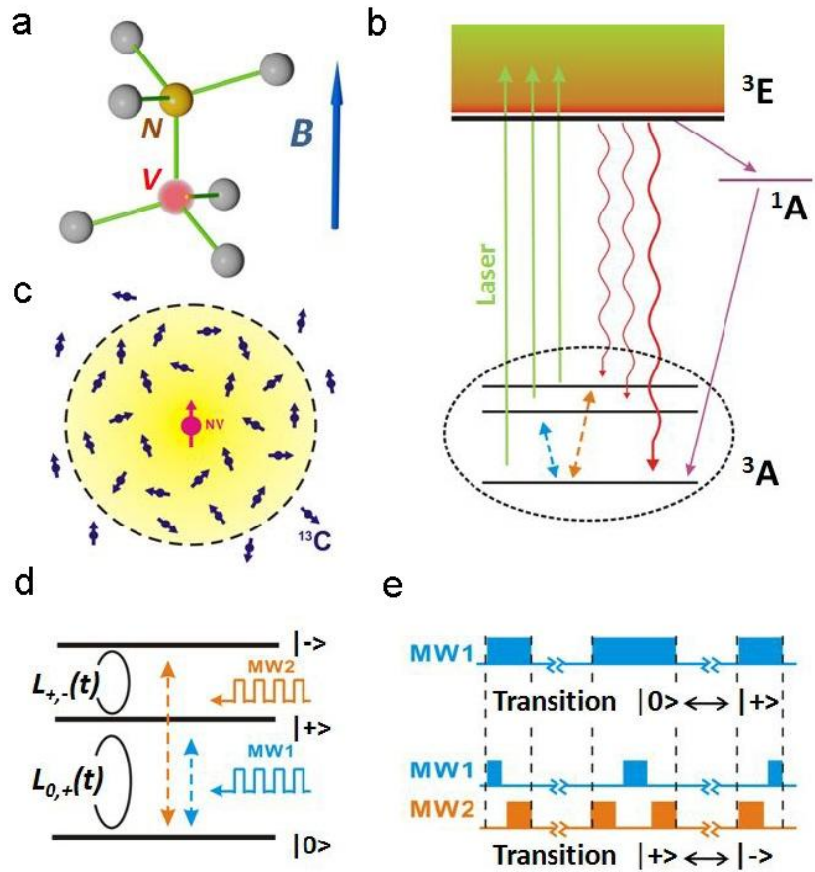

**Figure 1 | System and methods for measuring NV centre spin decoherence in the $^{13}$C nuclear spin bath in diamond. a**, Atomic structure of an NV centre in diamond and the magnetic field **B**. **b**, Energy levels and optical transitions for optical pump and detection of an NV centre spin. **c**, Schematic of a $^{13}$C nuclear spin bath (enclosed by the circle), which together with the NV centre spin form a coupled close quantum system. **d**, The single-transition coherence $L_{0,+}$ and the double-transition coherence $L_{+,-}$ between the NV centre triplet states. **e**, Microwave pulse sequences for controlling the centre spin. The double-transition is controlled by composite pulses in resonance with the single-transitions $|0\rangle \leftrightarrow |+\rangle$ (blue channel) and $|0\rangle \leftrightarrow |-\rangle$ (orange channel). In the single-transition coherence measurement, the $\pi/2$ and $\pi$ rotations are replaced with $3\pi/2$ and $3\pi$ rotations, respectively, to make the pulse durations the same as those used in the double-transition control. The typical duration of a $\pi$ rotation is 90 ns.

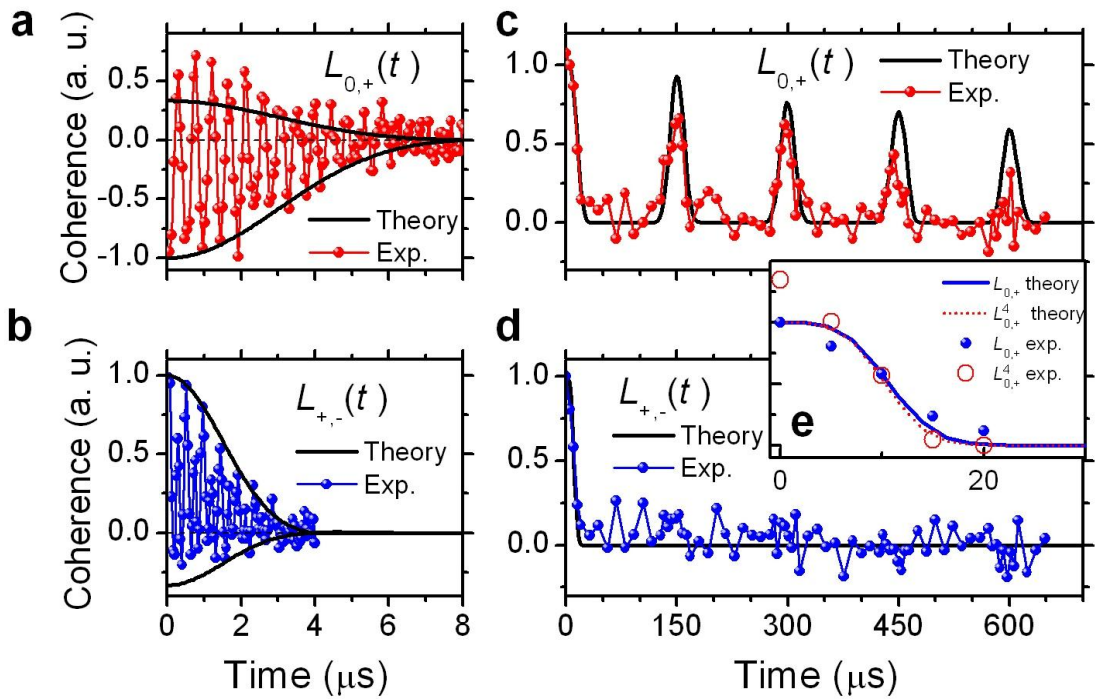

**Figure 2 | Free-induction decay and Hahn echo of the NV centre spin coherence. a** & **b**, Measured (color lines with symbols) and calculated (envelopes in black lines) free-induction decay of the single- and double-transition coherence, respectively. The oscillations and the asymmetric envelopes are due to coupling to the $^{14}$N nuclear spin. **c** & **d**, Measured (color lines with symbols) and calculated (black lines) Hahn-echo signals of the single- and the double-transition, respectively. **e**, Close-up of the initial decay in **c** & **d**, with the single-transition coherence scaled by the fourth power for comparison. A magnetic field $B = 12.5$ Gauss is applied along the NV axis.

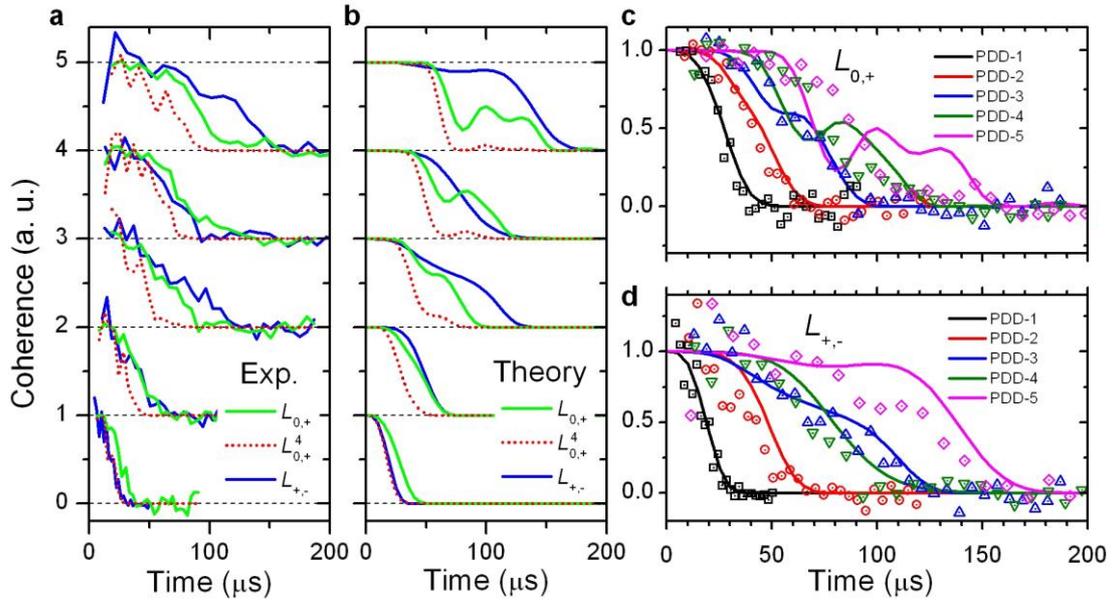

**Figure 3 | Decoherence of the NV centre spin under PDD control. a**, Measured single- and double-transition coherence, under the one- to five-pulse PDD control (PDD-1 to PDD-5, from bottom to top, vertically shifted for the sake of clarity). The scaled single-transition coherence $L_{0,+}^{4}$ is plotted for comparison. **b**, The calculated data, plotted in the same format as in **a**. **c**, Comparison between the measured (symbols) and the calculated (solid lines) single-transition coherence. **d**, The same as **c**, but for the double-transition. A magnetic field $B = 5$ Gauss is applied along the NV axis.

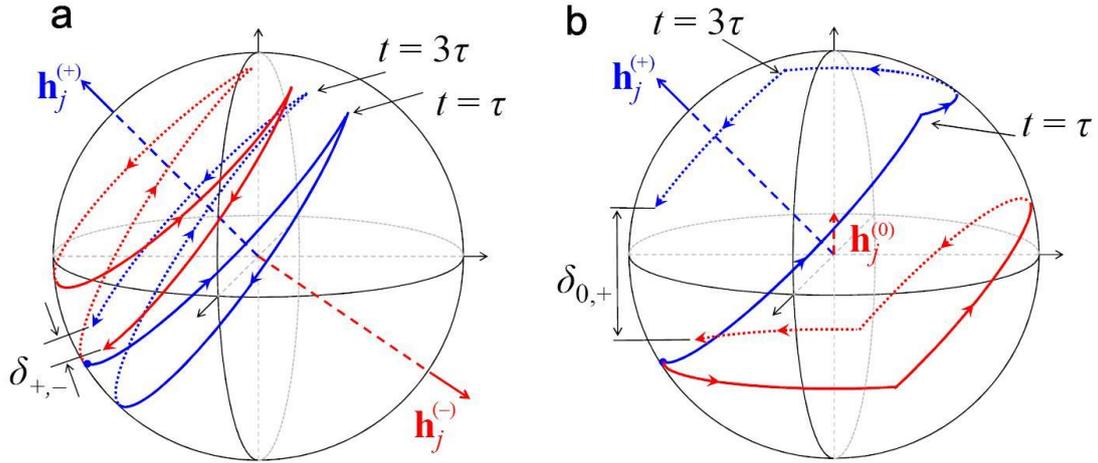

**Figure 4 | Nuclear spin precession conditioned on the centre spin state. a**, Bifurcated nuclear spin precession about the local fields $\mathbf{h}_j^{(+)}$ (blue arrow) and $\mathbf{h}_j^{(-)}$ (red arrow) under the 2-pulse PDD control of the double-transition. The blue (red) path shows the nuclear spin precessing about $\mathbf{h}_j^{(+)}$ ($\mathbf{h}_j^{(-)}$) from time 0 to $\tau$ for the centre spin state $|+\rangle$ ($|-\rangle$), then precessing about $\mathbf{h}_j^{(-)}$ ($\mathbf{h}_j^{(+)}$) from time $\tau$ to $3\tau$ after the centre spin is flipped to $|-\rangle$ ($|+\rangle$) at time $\tau$, and then precessing about $\mathbf{h}_j^{(+)}$ ($\mathbf{h}_j^{(-)}$) from time $3\tau$ to $4\tau$ after the centre spin is flipped back to $|+\rangle$ ($|-\rangle$) at time $3\tau$. **b**, The same as **a**, but for the single-transition. $\delta_{+,-}$ and $\delta_{0,+}$ denote the distances between the bifurcated paths at the echo time ($4\tau$) for the double- and single-transitions, respectively. As the two local fields $\mathbf{h}_j^{(\pm)}$ are almost anti-parallel, $\delta_{+,-}$ is much smaller than $\delta_{0,+}$.